\documentstyle[12pt]{article}
\begin{document}
\begin{center}
{\Large \bf
Decay of proton into Planck neutrino in the theory of gravity}
\bigskip

{\large D.L.~Khokhlov}
\smallskip

{\it Sumy State University, R.-Korsakov St. 2\\
Sumy 40007 Ukraine\\
e-mail: khokhlov@cafe.sumy.ua}
\end{center}

\begin{abstract}
It is considered gravitational interaction within the framework
of the Newton theory and the quantum field theory.
It is introduced the Planck neutrino $\nu_{Pl}$.
Gravitational interaction of the fields $\psi\psi$ includes
short-range interaction $\psi\nu_{Pl}$ and
long-range interaction $\nu_{Pl}\nu_{Pl}$.
Gravitational radiation can be identified with
the Planck neutrino.
The theory predicts the decay of proton into Planck neutrino.
It is assumed
that the Planck mass built from three fundamental constants
$\hbar$, $c$ and $G$ is fixed in all the inertial frames.
This leads to that the lifetime of proton relative to
the decay into Planck neutrino
decreases with the Lorentz factor as $\sim \gamma^{-5}$.
Such a dependence of the lifetime of proton
on the Lorentz factor yields a cut-off in the EHECRs spectrum.
It is shown that the first "knee" in the EHECRs spectrum
$E\sim 3\times 10^{15}\ {\rm eV}$
corresponds to the lifetime of proton equal to the lifetime
of the universe,
the second "knee" $E\sim 10^{17}-10^{18}\ {\rm eV}$
corresponds to the lifetime of proton equal to the
thickness of our galactic disc. The EHECRs with the energies
$E>3 \times 10^{18}\ {\rm eV}$ can be identified with
the Planck neutrinos.
\end{abstract}

\section{Introduction}

The principle of equivalence of inertial and gravitational masses
underlines the theory of gravity.
In the Einstein theory of gravity~\cite{M},
this leads to that the free gravitational field is nonlocalized.
Under the presence of the matter, the gravity is described by
the Einstein equations
\begin{equation}
G_{ik}=T_{ik}
\label{eq:GT}
\end{equation}
where $G_{ik}$ is the Einstein tenzor,
$T_{ik}$ is the tenzor of momentum-energy of the matter.
Free gravitational field defined by the absence of the matter
$T_{ik}=0$ is described by the equations
\begin{equation}
R_{ik}=0
\label{eq:R}
\end{equation}
where $R_{ik}$ is the Ricci tenzor.
The localized field must be described by the tenzor of momentum-energy.
Einstein characterized the momentum-energy of the
gravitational field by the pseudo-tenzor defined as
\begin{equation}
t^{ik}=H^{ilkm}_{\ \ \quad ,lm} -G^{ik}
\label{eq:tik}
\end{equation}
where $H^{ilkm}_{\ \ \quad ,lm}$ is the linearized part of $G_{ik}$.
Thus in the Einstein theory gravitational field is nonlocalized.

The natural way
proposed by Lorentz and Levi-Civita~\cite{Pau} is to take
$G_{ik}$ as the momentum-energy of the gravitational field. 
However in this
case $G_{ik}$ is equal to zero for the free gravitational field
$G_{ik}=R_{ik}=0$. Such a situation may be interpreted as that
the gravitational interaction occurs without gravitational field.
Then the problem arises as to how to introduce gravitational
radiation. The possible resolution of the problem is to introduce
some material field as a radiation.

\section{Theory}

Consider gravitational interaction within the framework of the Newton
theory and the quantum field theory.
The Lagrangian of the Newton gravity is given by
\begin{equation}
L=G\frac{m^2}{r},
\label{eq:L1}
\end{equation}
with the mass m being the gravitational charge.
While expressing the Newton constant $G$
via the charge $g=(\hbar c)^{1/2}$
and via the Planck mass $m_{Pl}=(\hbar c/G)^{1/2}$,
the Lagrangian (\ref{eq:L1}) can be rewritten in the form
\begin{equation}
L=G\frac{m^2}{r}=\frac{g^2}{m_{Pl}^2}\frac{m^2}{r}.
\label{eq:L2}
\end{equation}
The Lagrangian of the Newton gravity in the form (\ref{eq:L2})
describes gravitational interaction by means of the charge $g$.
In this way gravity may be implemented into
the quantum field theory.

Rewrite the Lagrangian (\ref{eq:L2}) in the form of the effective
Lagrangian of interaction of the spinor fields~\cite{B}
\begin{equation}
L=\frac{g^2}{m_{Pl}^2}J_{\mu}(x)J^{\mu}(x).
\label{eq:L5}
\end{equation}
The term $1/m_{Pl}^2$ in the Lagrangian (\ref{eq:L5}) reads that
gravitational interaction takes place at the Planck scale.
At the same time gravity is characterized by the infinite radius
of interaction.
To resolve the problem consider the scheme
of gravitational interaction which includes both the
short-range interaction and the long-range interaction
\begin{equation}
L=L_{short}+L_{long}.
\label{eq:L3}
\end{equation}
Let us introduce the Planck neutrino $\nu_{Pl}$.
Let the Planck neutrino is the massless particle of the spin 1/2.
Suppose that
the Planck neutrino interacts with the other fields at the Planck scale
\begin{equation}
\psi\rightarrow \nu_{Pl}
\label{eq:psnu}
\end{equation}
where $\psi$ denotes all the fields of the spin 1/2.
This interaction is of short-range and is governed
by the Lagrangian (\ref{eq:L5})
\begin{equation}
L_{short}=\frac{g^2}{m_{Pl}^2}J_{\mu}(x)J^{\mu}(x)
\label{eq:L6}
\end{equation}
where the current
$J_{\mu}$ transforms the field $\psi$ into the field $\nu_{Pl}$.
Let the interaction of the Planck neutrinos $\nu_{Pl}\nu_{Pl}$
is of long-range and
is governed by the Lagrangian identically equal to zero
\begin{equation}
L_{long}\equiv 0.
\label{eq:L4}
\end{equation}
The considered scheme allows one to decribe both the classical
gravity and the decay of the field $\psi$ into the Planck neutrino.
In this scheme gravitational radiation can be identified with
the Planck neutrino.

Within the framework of the standard quantum field theory,
the above scheme of gravitational interaction should include
two intermediate fields $\psi\nu_{Pl}$ and $\nu_{Pl}\nu_{Pl}$.
Since the Lagrangian of the interaction $\nu_{Pl}\nu_{Pl}$ is
identically equal to zero,
the energy of the field $\nu_{Pl}\nu_{Pl}$
is identically equal to zero.
The field $\psi\nu_{Pl}$ is defined by the Planck mass.
In the theory of gravity there is
the limit of ability to measure the length equal to
the Planck length~\cite{Tr}
$\Delta l\geq 2(\hbar G/c^3)^{1/2}=2l_{Pl}$.
From this it follows that there is no possibility to measure the
field $\psi\nu_{Pl}$ in the physical experiment.
Thus both intermediate fields
$\psi\nu_{Pl}$ and $\nu_{Pl}\nu_{Pl}$ cannot be measured.
This means that the intermediate fields $\psi\nu_{Pl}$
and $\nu_{Pl}\nu_{Pl}$ do not exist.
We arrive at the conclusion that gravitational interaction
occurs without intermediate fields.

\section{The lifetime of proton relative to the decay
into Planck neutrino}

In view of eq.~(\ref{eq:psnu}),
the decay of proton into Planck neutrino occurs at the Planck scale
\begin{equation}
p\rightarrow \nu_{Pl}.
\label{eq:pnu}
\end{equation}
The lifetime of proton relative to the decay
into Planck neutrino is defined by the Lagrangian (\ref{eq:L5})
\begin{equation}
t_p=t_{Pl}\left(\frac{m_{Pl}}{2m_p}\right)^5
\label{eq:tp1}
\end{equation}
where the factor 2 takes into account the transition from
the massive particle to the massless one.
This lifetime corresponds to the rest frame.
Consider the lifetime of proton in the moving frame with the
Lorentz factor
\begin{equation}
\gamma=\left(1-\frac{v^2}{c^2}\right)^{-1/2}.
\label{eq:gam}
\end{equation}
In the moving frame the rest mass and time are multiplied by the
Lorentz factor
\begin{equation}
m'=\gamma m
\label{eq:gm}
\end{equation}
\begin{equation}
t'=\gamma t.
\label{eq:gt}
\end{equation}
The Planck mass $m_{Pl}=(\hbar c/G)^{1/2}$ and the Planck time
$t_{Pl}=(\hbar G/c^5)^{1/2}$ are built from three fundamental
constants $\hbar$, $c$ and $G$.
According to the special relativity~\cite{Pau}, the speed of light
is fixed in all the inertial frames.
Extend the special relativity principle and suppose
that the three constants $\hbar$, $c$ and $G$ are fixed in all the
inertial frames
\begin{equation}
\hbar'=\hbar \qquad c'=c \qquad G'=G.
\label{eq:thr}
\end{equation}
Hence the Planck mass and time are fixed in all the
inertial frames.
Then the lifetime of proton in the moving frame
is given by
\begin{equation}
t_p '=t_{Pl}\left(\frac{m_{Pl}}{2\gamma m_p}\right)^5.
\label{eq:tp2}
\end{equation}

For comparison consider the decay of muon which is governed by
the Lagrangian of electroweak interaction~\cite{B}
\begin{equation}
L=\frac{g^2}{m_{W}^2}J_{\mu}(x)J^{\mu}(x)
\label{eq:Lmu}
\end{equation}
where $m_{W}$ is the mass of W-boson.
In the rest frame the lifetime of muon
is given by
\begin{equation}
t_{\mu}=t_{W}\left(\frac{m_{W}}{m_{\mu}}\right)^5.
\label{eq:tmu1}
\end{equation}
In the moving frame the lifetime of muon
is given by
\begin{equation}
t_{\mu}'=\gamma t_{W}\left(\frac{m_{W}}{m_{\mu}}\right)^5.
\label{eq:tmu2}
\end{equation}

Thus unlike the usual situation when the lifetime of the particle,
e. g. muon, grows with the Lorentz factor
as $\sim \gamma$,
the lifetime of proton relative to the decay into Planck neutrino
decreases with the Lorentz factor
as $\sim \gamma^{-5}$. State once again that such a behaviour
is due to that the Planck mass built from three fundamental constants
$\hbar$, $c$ and $G$ is fixed in all the inertial frames.

\section{Extra high energy cosmic rays spectrum
in view of the decay of proton into Planck neutrino}

In view of eq.~(\ref{eq:tp2}),
the lifetime of proton relative to the decay into Planck neutrino
decreases with the increase of the kinetic energy of proton.
Then the decay of proton can be observed for the extra high energy
protons. In particular the decay of proton can be observed as a cut-off
in the energy spectrum of extra high energy cosmic rays (EHECRs).

The EHECRs spectrum above $10^{10}\ {\rm eV}$ can be
divided into three regions: two "knees" and one "ankle"~\cite{Yo}.
The first "knee" appears around $3\times 10^{15}\ {\rm eV}$
where the spectral power law index changes from $-2.7$ to $-3.0$.
The second "knee" is somewhere between $10^{17}\ {\rm eV}$ and
$10^{18}\ {\rm eV}$ where the spectral slope steepens from
$-3.0$ to around $-3.3$. The "ankle" is seen in the region of
$3 \times 10^{18}\ {\rm eV}$ above which the spectral slope
flattens out to about $-2.7$.

Consider the EHECRs spectrum
in view of the decay of proton into Planck neutrino.
Let the earth be the rest frame.
For protons arrived at the earth,
the travel time meets the condition
\begin{equation}
t\leq t_p.
\label{eq:t}
\end{equation}
From this the time required for proton travel from the source to the
earth defines the limiting energy of proton
\begin{equation}
E_{lim}=\frac{m_{Pl}}{2}\left(\frac{t_{Pl}}{t}\right)^{1/5}.
\label{eq:E}
\end{equation}
Within the time $t$, protons
with the energies $E>E_{lim}$ decay and
do not give contribution in the EHECRs spectrum.
Thus the energy $E_{lim}$ defines a cut-off in the
EHE proton spectrum. Planck neutrinos appeared due to the
decay of the EHE protons may give 
a contribution in the EHECRs spectrum. 
If the contribution of Planck neutrinos
in the EHECRs spectrum is less compared with the contribution
of protons one can observe the cut-off at the energy $E_{lim}$
in the EHECRs spectrum.

Determine
the range of the limiting energies of proton
depending on the range of distances to the EHECRs sources.
Take the maximum and minimum distances to the source as
the size of the universe and the thickness
of our galactic disc respectively.
For the lifetime of the universe
$\tau_0=14 \pm 2 \ {\rm Gyr}$~\cite{age},
the limiting energy is equal to $E_1=3.9 \times 10^{15}\ {\rm eV}$.
This corresponds to the first "knee" in the EHECRs spectrum.
For the thickness of our galactic disc $\simeq 300\ {\rm pc}$,
the limiting energy is equal to $E_2=5.5 \times 10^{17}\ {\rm eV}$.
This corresponds to the second "knee" in the EHECRs spectrum.
Thus
the range of the limiting energies of proton
due to the decay of proton into Planck neutrino
lies between
the first "knee" $E\sim 3\times 10^{15}\ {\rm eV}$ and
the second "knee" $E\sim 10^{17}-10^{18}\ {\rm eV}$.

From the above consideration it follows that
the decrease of the spectral power law index from $-2.7$ to $-3.0$
at the first "knee" $E\sim 3\times 10^{15}\ {\rm eV}$ and
from $-3.0$ to around $-3.3$
at the second "knee" $E\sim 10^{17}-10^{18}\ {\rm eV}$
can be explained as a result of
the decay of proton into Planck neutrino.
From this it seems natural that, below
the "ankle" $E<3 \times 10^{18}\ {\rm eV}$,
the EHECRs events are mainly caused by the protons.
Above the "ankle" $E>3 \times 10^{18}\ {\rm eV}$,
the EHECRs events are caused by the particles other than protons.

If Planck neutrinos take part in the strong interactions,
they must give some contribution in the EHECRs events.
To explain the observed EHECRs spectrum
it is necessary to assume that the contribution of Planck neutrinos
in the EHECRs spectrum is less compared with the contribution
of protons.
Suppose that proton decays into 5 Planck neutrinos.
Then the energy of the Planck neutrino is $1/5$ of the energy of
the decayed proton.
For the spectral power law index equal to $-2.7$,
the ratio of the proton flux to the Planck neutrino flux
is given by
$J_p/J_{\nu}=5^{1.7}=15.4$.

From the above consideration it is natural to identify EHE
particles with the energies $E>3 \times 10^{18}\ {\rm eV}$
with the Planck neutrinos.
Continue the curve with the spectral power law index $-2.7$
from the "ankle" $E\sim 3 \times 10^{18}\ {\rm eV}$ to
the first "knee" $E\sim 3\times 10^{15}\ {\rm eV}$ and
compare the continued curve with the observational curve.
Comparison gives
the ratio of the proton flux to the Planck neutrino flux
$J_p/J_{\nu}\approx 15$.

\end{document}